\documentstyle[12pt]{article}
\input psfig.sty
\def\nn{\noindent}
\def\Re{{\cal R \mskip-4mu \lower.1ex \hbox{\it e}\,}}
\def\Im{{\cal I \mskip-5mu \lower.1ex \hbox{\it m}\,}}
\def\ie{{\it i.e.}}
\def\eg{{\it e.g.}}

\def\etal{{\it et al.}}

\def\sub#1{_{\lower.25ex\hbox{$\scriptstyle#1$}}}
\def\tev{\,{\ifmmode\mathrm {TeV}\else TeV\fi}}
\def\gev{\,{\ifmmode\mathrm {GeV}\else GeV\fi}}
\def\mev{\,{\ifmmode\mathrm {MeV}\else MeV\fi}}
\def\to{\rightarrow}

\def\subw{_{\rm w}}
\def\mh{\ifmmode m\sbl H \else $m\sbl H$\fi}
\def\mch{\ifmmode m_{H^\pm} \else $m_{H^\pm}$\fi}
\def\mt{\ifmmode m_t\else $m_t$\fi}
\def\mc{\ifmmode m_c\else $m_c$\fi}
\def\mz{\ifmmode M_Z\else $M_Z$\fi}
\def\mw{\ifmmode M_W\else $M_W$\fi}
\def\mws{\ifmmode M_W^2 \else $M_W^2$\fi}
\def\mhs{\ifmmode m_H^2 \else $m_H^2$\fi}   
\def\mzs{\ifmmode M_Z^2 \else $M_Z^2$\fi}
\def\mts{\ifmmode m_t^2 \else $m_t^2$\fi}
\def\mcs{\ifmmode m_c^2 \else $m_c^2$\fi}
\def\mchs{\ifmmode m_{H^\pm}^2 \else $m_{H^\pm}^2$\fi}
\def\ztwo{\ifmmode Z_2\else $Z_2$\fi}
\def\zone{\ifmmode Z_1\else $Z_1$\fi}
\def\mtwo{\ifmmode M_2\else $M_2$\fi}
\def\mone{\ifmmode M_1\else $M_1$\fi}
\def\tb{\ifmmode \tan\beta \else $\tan\beta$\fi}
\def\xw{\ifmmode x\subw\else $x\subw$\fi}
\def\ch{\ifmmode H^\pm \else $H^\pm$\fi}
\def\lum{\ifmmode {\cal L}\else ${\cal L}$\fi}
\def\inpb{\,{\ifmmode {\mathrm {pb}}^{-1}\else ${\mathrm {pb}}^{-1}$\fi}}
\def\infb{\,{\ifmmode {\mathrm {fb}}^{-1}\else ${\mathrm {fb}}^{-1}$\fi}}
\def\epem{\ifmmode e^+e^-\else $e^+e^-$\fi}
\def\ppb{\ifmmode \bar pp\else $\bar pp$\fi}
\def\bsg{\ifmmode B\to X_s\gamma\else $B\to X_s\gamma$\fi}
\def\bsll{\ifmmode B\to X_s\ell^+\ell^-\else $B\to X_s\ell^+\ell^-$\fi}
\def\bstt{\ifmmode B\to X_s\tau^+\tau^-\else $B\to X_s\tau^+\tau^-$\fi}
\def\lamt{\ifmmode \tilde\lambda\else $\tilde\lambda$\fi}
\def\shat{\ifmmode \hat s\else $\hat s$\fi}
\def\that{\ifmmode \hat t\else $\hat t$\fi}
\def\uhat{\ifmmode \hat u\else $\hat u$\fi}

\newskip\zatskip \zatskip=0pt plus0pt minus0pt
\def\matth{\mathsurround=0pt}

\def\atversim#1#2{\lower0.7ex\vbox{\baselineskip\zatskip\lineskip\zatskip
  \lineskiplimit 0pt\ialign{$\matth#1\hfil##\hfil$\crcr#2\crcr\sim\crcr}}}

\renewcommand{\thefootnote}{\fnsymbol{footnote}}

\begin{document} \begin{titlepage} 
\rightline{\vbox{\halign{&#\hfil\cr
&SLAC-PUB-7982\cr
&November 1998\cr}}}
\begin{center}

{\Large\bf Distinguishing Indirect Signatures of New Physics at the NLC: 
$Z^\prime$ Versus $R$-Parity Violation}
\footnote{Work supported by the Department of 
Energy, Contract DE-AC03-76SF00515}
\medskip

\normalsize 
{\large Thomas G. Rizzo } \\
\vskip .3cm
Stanford Linear Accelerator Center \\
Stanford University \\
Stanford CA 94309, USA\\
\vskip .3cm

\end{center}

\begin{abstract} 
$R$-parity violation and extensions of the Standard Model gauge structure 
offer two non-minimal realizations of supersymmetry at low energies that can 
lead to similar new physics signatures at existing and future colliders. We 
discuss techniques that can be employed at the NLC below direct production 
threshold to distinguish these two new physics scenarios. 
\end{abstract} 




\renewcommand{\thefootnote}{\arabic{footnote}} \end{titlepage}


\section{Introduction}

While the Standard Model(SM) is in relatively good agreement with all precision 
electroweak data{\cite {rev}}, it leaves too many unanswered questions that 
will somehow need to be addressed by new physics at or above the electroweak 
scale. Supersymmetry(SUSY), in 
the guise of the Minimal Supersymmetric Standard Model(MSSM), provides a 
potential starting point for the exploration of this new physics; however,  
while the MSSM provides a simplified framework in which to work, most authors 
would agree that the MSSM is itself inadequate due to the very large 
number of free parameters it contains. Furthermore, the MSSM cannot be the 
whole story of low-energy SUSY since, on its own, it does not explain how 
SUSY is broken or why the scale of this breaking is of order $\sim 1$ TeV. 
In going beyond the MSSM there are many possible paths to follow. In this 
paper we discuss two of the simplest of these scenarios: an extension of the 
SM gauge group by an additional $U(1)$ factor broken near the TeV scale and 
$R$-parity violation, both of which are well-motivated by string theory.  
Although these two alternatives would appear to have little in common, we will 
see below that they can lead to similar phenomenology at present and future 
colliders and may be easily confused in certain regions of 
the parameter space for each class of model.

Unlike the case of Grand Unified Theories(GUT), where any additional $U(1)$'s 
may break at any arbitrary scale below $M_{GUT}$, perturbative string models 
with gravity 
mediated SUSY breaking are known to predict an assortment of new gauge bosons 
with masses of order 1 TeV, 
as well as the existence of other exotic matter states with comparable 
masses{\cite {cv}}. Such models lead one to expect that the existence of a 
$Z'$ at mass scales which will be accessible at Run II of the Tevatron or at 
future colliders is quite natural. Similarly, the case for potential 
$R$-parity violation is also easily demonstrated and appears to be just as 
natural as not. As is well known, the conventional 
gauge symmetries of the supersymmetric extension of the 
SM allow for the existence of additional terms in the superpotential that 
violate either Baryon($B$) and/or Lepton($L$) number. One quickly 
realizes that simultaneous existence of such terms leads to rapid proton 
decay. These phenomenologically dangerous terms can be written as
\begin{equation}
W_R=\lambda_{ijk}L_iL_jE^c_k+\lambda'_{ijk}L_iQ_jD^c_k+\lambda''_{ijk}
U^c_iD^c_jD^c_k+\epsilon_iL_iH\,, 
\end{equation}
where $i,j,k$ are family indices and symmetry demands that $i<j(j<k)$ in the 
terms proportional to $\lambda$($\lambda''$) Yukawa couplings. In the 
MSSM, the imposition of the discrete symmetry of $R$-parity removes by brute 
force all of these `undesirable' 
couplings from the superpotential. However, it easy to construct alternative 
discrete symmetries which may arise from strings that allow for the 
existence of either the $L$- or 
$B$-violating terms~{\cite {herbi}} in $W_R$ (but not both kinds) and are just 
as likely to exist as $R$-parity itself. 
(Interestingly, at least some, if not all, of these dangerous couplings in 
$W_R$ may be removed from the superpotential if the SM 
fields also carry an additional set of $U(1)$ quantum numbers{\cite {old}}.)
As far as we know there exists no strong 
theoretical reason to favor the MSSM realization over such $R$-parity 
violating scenarios. 
Since only $B$- or $L$-violating terms survive when this new symmetry is 
present the proton now remains stable in these models. Consequently, 
various low-energy phenomena then provide the only 
significant constraints~{\cite {constraints}} on the Yukawa couplings 
$\lambda,\lambda'$ and $\lambda''$. For example, constraints on the 
trilinear $LLE^c$ 
couplings are typically of order $\lambda \sim 0.05 (m/100 GeV)$, 
where $m$ 
is the mass of the exchanged sfermion. In what follows we will be interested 
in $\tilde \nu$ masses in the TeV range so that Yukawa couplings not much less 
than unity can be phenomenologically viable. 

If $R$-parity is violated much of the conventional wisdom associated with the 
phenomenology of the MSSM goes by the 
wayside, \eg, the LSP (now not necessarily a neutralino!) is 
unstable and sparticles may now be produced singly. In particular, it is 
possible that the exchange of sparticles can significantly modify SM processes 
and may even be produced as $s$-channel resonances, 
appearing as bumps~{\cite {bumps,bumps2}} in cross sections if 
they are kinematically 
accessible. Below threshold, these new spin-0 exchanges may make 
their presence known via indirect effects on cross sections and 
other observables even when they occur in the 
$t$- or $u$-channels{\cite {tandj}}. Here we will address the 
question of whether the 
effects of the exchange of such particles can be differentiated from those 
conventionally associated with a $Z'$. (Recall the expectation that at 
linear colliders such as the NLC, the effects of a $Z'$ with a mass in the 
several TeV range will appear as deviations from the SM values for observables 
associated with the processes $e^+e^-\to f\bar f$.)

In many cases it will be quite straightforward to differentiate these two 
alternative sources of new physics. For example, if a new resonance is 
actually produced and is found to 
dominantly decay to SUSY partners, including gauginos,  
or violate lepton number, we will know immediately that the new particle is  
most probably a sfermion with couplings that result 
from $R$-parity violation. If, on the otherhand, such a 
particle were to be produced at a lepton or hadron collider and dominantly 
decay to SM fields, the angular distribution of the final state products, 
either leptons or jets, would conclusively tell us{\cite {bumps}} the spin 
of the resonance given sufficient statistics, \ie, several hundred events. 
We will not be concerned with this scenario below. 

The situation becomes far more uncertain, however, when below threshold 
exchanges are involved and the existence of the interaction produced by the 
new particle is uncovered {\it only} through its modification of 
cross sections and asymmetries for SM processes. 
As an example, both a leptophobic $Z'$ and a squark 
coupling via the $B$-violating $U^cD^cD^c$ term in $W_R$ can alter the 
angular distribution of dijets via an $s$-channel exchange below threshold 
at the Tevatron. It is {\it not} so obvious that these two scenarios can be 
easily, if at all, distinguished by a detailed analysis of these deviations.  

Since we are concerned here with NLC physics we will by necessity limit our 
attention solely to the trilinear 
$L$-violating terms in the superpotential. If only the 
$LLE^c$ terms are present it is clear that only the observables associated with 
leptonic processes will be affected by the exchange of $\tilde \nu$'s in the 
$s$- or $t$-channels or both 
and no input into the analysis from hadron collider experiments is 
possible. On the otherhand, if $LQD^c$ terms are also present then the 
$Q=-1/3$ final states at linear colliders will also potentially be affected 
by $\tilde \nu$ exchange. 
Simultaneously a $\tilde \nu$ resonance may show up at a hadron collider 
in the Drell-Yan or dijet channels if 
kinematically allowed and the Yukawa couplings to first generation down-type 
quarks is sizeable. In the analysis below we will consider for simplicity only 
the former situation; the extension of our analysis to the more general case 
involving final state quarks 
is quite straightforward. This implies that we will be directly comparing the 
$s$-channel exchange of an essentially 
hadrophobic $Z'$ with $\tilde \nu$ exchanges. 

How does a generic $Z'$ couple to leptons? In most GUT-type 
models, $Z'$ couplings are 
both flavor diagonal and universal, \ie, generation independent. However, it 
is easy to construct more generalized models{\cite {nonunz}} 
where the $Z'$ couplings remain flavor diagonal 
but are rendered generation-dependent. It is this specific class of $Z'$ 
models 
which we will consider below since they mimic $\tilde \nu$'s most closely. 
Thus, while observing {\it different} deviations in the 
$e^+e^-\to e^+e^-,\mu^+\mu^-$ and $\tau^+\tau^-$ processes might be considered 
a unique $R$-parity violating signature, we see here that this need not be 
generally true, \ie, universality violation is not necessarily a smoking gun 
signal for $R$-parity violation.

\begin{table*}[htbp]
\leavevmode
\begin{center}
\label{snus1}
\begin{tabular}{lcc}
\hline
\hline
Reaction & Yukawa Coupling & Exchange(s) \\
\hline
$e^+e^- \to e^+e^-$  &$\lambda_{121}$   & $\tilde \nu_{\mu}(s,t)$   \\
                     &$\lambda_{131}$   & $\tilde \nu_{\tau}(s,t)$   \\
\hline
$e^+e^- \to \mu^+\mu^-$  &$\lambda_{121}$  & $\tilde \nu_{e}(t)$   \\
                         &$\lambda_{122}$  & $\tilde \nu_{\mu}(t)$   \\
                         &$\lambda_{132}$  & $\tilde \nu_{\tau}(t)$   \\
                         &$\lambda_{231}$  & $\tilde \nu_{\tau}(t)$   \\
\hline
$e^+e^- \to \tau^+ \tau^-$  &$\lambda_{123}$   & $\tilde \nu_{\mu}(t)$   \\
                            &$\lambda_{131}$   & $\tilde \nu_{e}(t)$   \\
                            &$\lambda_{133}$   & $\tilde \nu_{\tau}(t)$   \\
                            &$\lambda_{231}$   & $\tilde \nu_{\mu}(t)$   \\
\hline
\hline
\end{tabular}
\caption{Reactions that can be mediated by $\tilde \nu$'s if only one Yukawa 
coupling in the $LLE^c$ term of the superpotential is large. The $s$ and/or $t$ 
in the right hand column labels the exchange channel.}
\end{center}
\end{table*}

The conventional approach in analyzing $R$-parity violating phenomenology is to 
consider the case where only one or two of the Yukawa couplings in $W_R$ can 
be significantly large at a time{\cite {constraints,bumps}}. 
If we follow this approach we can 
immediately write down which reactions are modified by $s$- or $t$-channel 
$\tilde \nu$ exchanges for a given non-zero $\lambda$ or pair of $\lambda$'s 
at the NLC. For simplicity, any small mass splittings between sneutrinos and 
anti-sneutrinos will be ignored{\cite {yuval}} in this analysis. 
In the case when only one non-zero Yukawa coupling is present, 
Table 1 informs us that $\tilde \nu$'s may contribute to either 
$e^+e^-\to \mu^+\mu^-$ or $\tau^+\tau^-$ via $t$-channel exchange while 
$e^+e^-\to e^+e^-$ receives both $s$- and $t$-channel contributions. Note 
that if the $\lambda_{121}$, $\lambda_{131}$ or $\lambda_{231}$ 
are non-zero, $\tilde \nu$ 
exchange of different flavors can contribute to deviations in 
more than one final state. Table 2 shows us 
that if two Yukawas are simultaneously large, most final 
states are lepton family number violating, \eg, $e^+e^-\to e^+\tau^-$. In 
such cases, the separation of the $Z'$ and $R$-parity violation scenarios 
would again be straightforward since it is very unlikely that a TeV mass 
$Z'$ would have large lepton family number violating couplings. However, we 
also see from this Table that if only the product of Yukawas  
$\lambda_{121}\lambda_{233}$ or $\lambda_{131}\lambda_{232}$ is non-zero 
then $s$-channel $\tilde\nu$ exchange would contribute to the $\tau^+\tau^-$ 
or $\mu^+\mu^-$ final state, respectively. Putting this together with the 
results of Table 1 we see that if either of these two products of Yukawa 
couplings is non-zero 
all possible leptonic final states may receive contributions from 
$R$-parity violating $\tilde \nu$ exchanges. We now turn to a study of these 
various cases.

\begin{table*}[htbp]
\leavevmode
\begin{center}
\label{snus2}
\begin{tabular}{lcc}
\hline
\hline
Yukawa Couplings & Final State & Exchange(s) \\
\hline
$\lambda_{121}\lambda_{122}$  &$e\mu$   & $\tilde \nu_{\mu}(s,t)$   \\
$\lambda_{121}\lambda_{123}$  &$e\tau$   & $\tilde \nu_{\mu}(s,t)$   \\
$\lambda_{121}\lambda_{231}$  &$e\tau$   & $\tilde \nu_{\mu}(s)$   \\
$\lambda_{121}\lambda_{232}$  &$\mu\tau$   & $\tilde \nu_{\mu}(s)$   \\
$\lambda_{121}\lambda_{233}$  &$\tau\tau$   & $\tilde \nu_{\mu}(s)$   \\
$\lambda_{122}\lambda_{123}$  &$\mu\tau$   & $\tilde \nu_{\mu}(t)$   \\
$\lambda_{131}\lambda_{132}$  &$e\mu$   & $\tilde \nu_{\tau}(s,t)$   \\
$\lambda_{131}\lambda_{133}$  &$e\tau$   & $\tilde \nu_{\tau}(s,t)$   \\
$\lambda_{131}\lambda_{231}$  &$e\mu$   & $\tilde \nu_{\tau}(s)$   \\
$\lambda_{131}\lambda_{232}$  &$\mu\mu$   & $\tilde \nu_{\tau}(s)$   \\
$\lambda_{131}\lambda_{233}$  &$\mu\tau$   & $\tilde \nu_{\tau}(s)$   \\
$\lambda_{132}\lambda_{133}$  &$\mu\tau$   & $\tilde \nu_{\tau}(t)$   \\
\hline
\hline
\end{tabular}
\caption{$e^+e^-$ final states that can result from $\tilde \nu$ exchange in 
the $s$- and/or $t$-channels if two Yukawa couplings in the $LLE^c$ term of the 
superpotential are simultaneously non-zero.}
\end{center}
\end{table*}

The organization of this paper is as follows. 
In Section 2 of this paper we consider the case where the $\tilde \nu$ is 
exchanged in the $t$-channel leading to modifications in the reactions 
$e^+e^-\to \mu^+\mu^-$ and/or $\tau^+\tau^-$. $s$-channel exchange is discussed 
in Section 3 and Bhabha scattering in Section 4. Our summary and conclusions 
can be found in Section 5. We note that although we have only considered the 
case of $R$-parity exchanges in the $s$- and/or $t$-channels in this paper the 
analysis we follow can be easily adapted to other possible scalar (or higher 
spin) exchanges.

\section{$t$-channel $\tilde \nu$ Exchange}

In this section we will compare and contrast the $s$-channel $Z'$ contribution 
to $e^+e^-\to \mu^+\mu^-$ or $\tau^+\tau^-$ 
with that of a $\tilde \nu$ in the $t$-channel. To be specific, in the 
numerical analysis that follows we will consider a 1 TeV NLC with an 
integrated luminosity of $150~fb^{-1}$. 
The extension to other colliders with 
different center of mass energies and integrated luminosities is 
straightforward and can be partially obtained through a simple scaling 
relations{\cite {leike}}. With this luminosity almost all errors will be 
statistically dominated.  
Following{\cite {bumps}} the notational conventions of 
Kalinowski \etal, the differential cross section for the process $e^+e^-\to 
f\bar f$, where $f=\mu$ or $\tau$, allowing for possible $t$-channel 
$\tilde \nu$ or $s$-channel $Z'$ exchange, can be written as 
\begin{equation}
{d\sigma\over {dz}}={\pi\alpha^2\over {8s}}\Big[(1+z)^2\{ {1+P\over {2}}
|f^s_{LR}|^2+{1-P\over {2}}|f^s_{RL}|^2\}+(1-z)^2\{ {1+P\over {2}}
|f^s_{LL}|^2+{1-P\over {2}}|f^s_{RR}|^2\}\Big]\,,
\end{equation}
where $z=\cos \theta$, the angle with respect to the $e^-$ beam and 
\begin{eqnarray}
f^s_{LR}&=&1+P_Z(g^e_L)^2\oplus P_{Z'}g^{e\prime}_Lg^{f\prime}_L
\oplus 0\,, \nonumber \\
f^s_{RL}&=&1+P_Z(g^e_R)^2\oplus P_{Z'}g^{e\prime}_Rg^{f\prime}_R
\oplus 0\,, \nonumber \\
f^s_{LL}&=&1+P_Zg^e_Lg^e_R\oplus P_{Z'}g^{e\prime}_Lg^{f\prime}_R
\oplus {1\over {2}}C_{\tilde \nu}P^t_{\tilde \nu}\,, \nonumber \\
f^s_{RR}&=&1+P_Zg^e_Rg^e_L\oplus P_{Z'}g^{e\prime}_Rg^{f\prime}_L
\oplus {1\over {2}}C_{\tilde \nu}P^t_{\tilde \nu}\,,
\end{eqnarray}
where $P_{Z,Z'}=s/(s-M_{Z,Z'}^2+iM_{Z,Z'}\Gamma_{Z,Z'})\simeq 
s/(s-M_{Z,Z'}^2)$ provides an adequate approximation when 
$M_Z^2 \ll s\ll M_{Z'}^2$, 
$P^t_{\tilde \nu}=s/(t-m_{\tilde \nu}^2)$ with $t=-s(1-z)/2$, $C_{\tilde \nu}=
\lambda^2/4\pi\alpha$, with $\lambda$ being the relevant Yukawa coupling from 
the superpotential, and the $Z$ and $Z'$ gauge couplings are normalized 
such that 
$g^e_L=c({-1\over {2}}+x)$ and $g^e_R=cx$ with $x=\sin^2 \theta_w$ 
and $c=\{\sqrt 2 G_FM_Z^2/\pi \alpha\}^{1/2}$. By `$\oplus$' in the equation 
above we mean that we may choose either term, \ie, the term after the first 
$\oplus$ corresponds to a potential $Z'$ contribution while that after the 
second $\oplus$ arises due to $t$-channel 
$\tilde \nu$ exchange. In addition, we note that the parameter $P$ in the 
expression above represents the polarization of the incoming electron beam, 
which we take 
to be $90\%$ in our analysis below (although it's specific value will not be 
too important as we will soon see). This single beam polarization allows us to 
construct a $z$-dependent Left-Right Asymmetry, $A_{LR}(z)$:
\begin{equation}
A_{LR}(z)={(1+z)^2\{|f^s_{LR}|^2-|f^s_{RL}|^2\}+(1-z)^2\{|f^s_{LL}|^2
-|f^s_{RR}|^2\}\over {(1+z)^2\{|f^s_{LR}|^2+|f^s_{RL}|^2\}+(1-z)^2\{
|f^s_{LL}|^2+|f^s_{RR}|^2\} }}\,. 
\end{equation}
\vspace*{-0.5cm}
\nn
\begin{figure}[htbp]
\centerline{
\psfig{figure=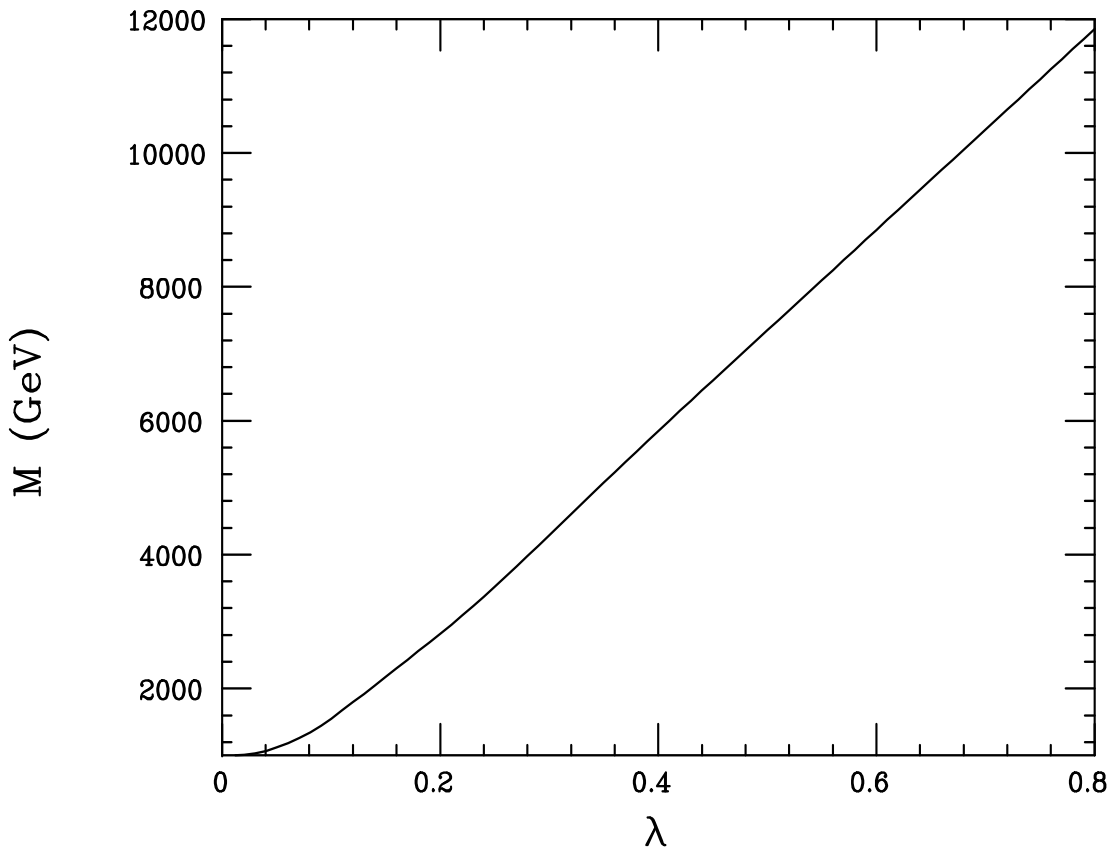,height=14cm,width=17cm,angle=-90}}
\vspace*{-1cm}
\caption[*]{{Indirect search reach for $t$-channel exchanged $\tilde \nu$'s 
as a function of their mass from the process $e^+e^-\to \mu^+\mu^-$ or 
$\tau^+\tau^-$ at a 1 TeV NLC with an integrated luminosity of 
\protect{$L=150~fb^{-1}$} including the effects of initial state radiation. The 
discovery region lies below the curve.}}
\label{fig1}
\end{figure}
\vspace*{0.4mm}

For a $Z'$ or $\tilde \nu$ with fixed couplings the first question one must 
address is the search reach for either particle 
assuming that only one of the $\mu^+\mu^-$ or $\tau^+\tau^-$ final states is 
affected. In the $Z'$ case, this result is essentially already 
documented{\cite {snow}}; for typical coupling strengths
the search reach for a $Z'$ is 
$(4.5-7)\sqrt s$ with the lower end of the range being the most relevant in our 
case due to the fact that only leptonic observables of a given flavor are now 
employed to set the limit. A similar analysis following an identical approach 
leads to Fig.1 which shows the corresponding reach for $\tilde \nu$ exchange 
in the $t$-channel. As in the $Z'$ case, for a fixed coupling strength 
we examine the deviations in the binned distributions for both the 
conventional production cross section as well as $A_{LR}(z)$ as functions of 
the $\tilde \nu$ mass accounting for both statistical and systematic errors 
after angular acceptance cuts of $10^o$ are imposed. Lepton identification 
efficiencies of $100\%$ are assumed for all three generations. The dominant 
systematic errors in the case of lepton final states are those associated 
with uncertainties in the machine luminosity and the beam polarization which 
we take from Ref.{\cite {snow}}. As we lower the $\tilde \nu$ mass from some 
initially very large value, the new physics effects become sufficiently large 
in comparison to the anticipated errors that the discovery of some type of 
new physics can be claimed. For more details of this procedure see 
Ref.{\cite {snow}}. It is important to 
remember that these search reaches are {\it only} telling us that new physics 
beyond the SM is definitely present but not what its nature may be. It is 
clear that only for a somewhat lighter $Z'$ or $\tilde \nu$ would sufficient 
statistics be available to differentiate the two new physics sources.

The angular distribution and $A_{LR}(z)$ provide us with potential tools to 
attack this problem.
Unfortunately, $A_{LR}(z)$ and/or the angular averaged quantity, $A_{LR}$, is 
numerically small at $\sqrt s=1 $ TeV and relatively poorly determined with 
integrated luminosities of $150~fb^{-1}$. For example, in the SM one finds 
$A_{LR}=(6.31\pm 1.06)\%$ assuming only statistical errors. In the numerical 
examples we will consider below, $A_{LR}$ is found to vary by no more 
than $\sim 0.5\sigma$ from this SM value and is thus not a good discriminator 
between $Z'$ and $\tilde \nu$ exchanges. This leaves us solely with the 
angular distribution with which to work and we will thus neglect the effects 
associated with single beam polarization in what follows. We note, however, 
that if $\tilde \nu$ exchange were to modify hadronic final states via the 
$LQD^c$ term in the superpotential we would find a significantly larger and 
much more useful value of $A_{LR}$ for those states.

\vspace*{-0.5cm}
\nn
\begin{figure}[htbp]
\centerline{
\psfig{figure=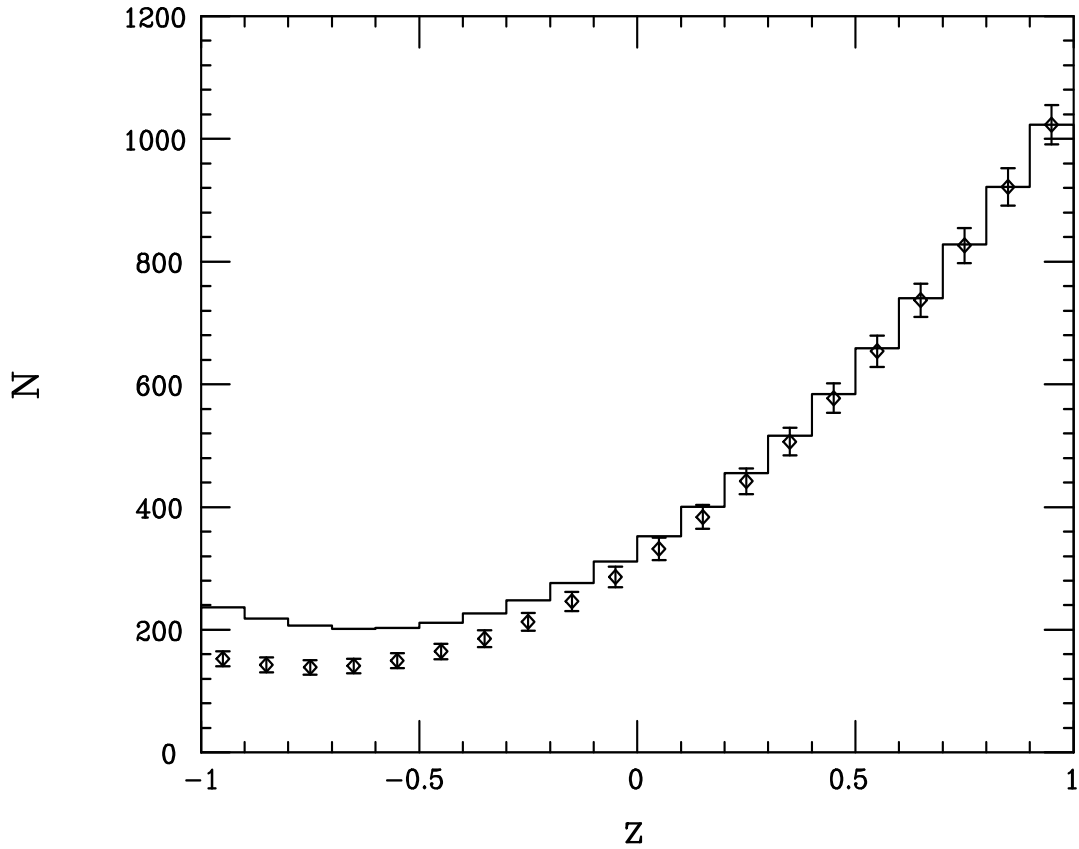,height=14cm,width=17cm,angle=-90}}
\vspace*{-1cm}
\caption[*]{Binned angular distribution for the process 
$e^+e^-\to \mu^+\mu^-$ or 
$\tau^+\tau^-$ at a 1 TeV NLC in the SM (histogram) and for the 
case where a 3 TeV $\tilde \nu$ with $\lambda=0.5$ exchanged in 
the $t$-channel also contributes. 
The errors are statistical only and represent an integrated luminosity of 
$L=150~fb^{-1}$. Initial state radiation has been included.}
\label{fig2}
\end{figure}
\vspace*{0.4mm}

At first glance one would think 
that these two new physics models are easily separable since the exchanges are 
in distinct channels. This is true provided we are reasonably sensitive to the 
$t$-dependent part of the $\tilde \nu$ propagator which would certainly not 
be the case if we were in the the contact interaction limit, \ie,  
$s,|t|\ll M^2_{Z',{\tilde \nu}}$. (As we will see below, 
this parameter space region is 
quite large.)  How does $Z'$ and $\tilde \nu$ exchange influence the angular 
distributions? Fig.2 shows the bin-integrated angular distribution for the 
$R$-parity violating case assuming $\lambda=0.5$ and a $\tilde \nu$ mass of 
3 TeV in comparison to that for the SM. Here we see the general feature that 
at large positive $z$ the two distributions completely agree 
but the $\tilde \nu$ 
exchange causes a depletion of events with negative $z$. We note from the 
figure that this depletion is clearly statistically meaningful. 
This result will hold 
for all interesting mass and coupling values and thus we learn that if an 
{\it increase} of the angular distribution is observed for negative $z$ the 
new physics that accounts for it {\it cannot} arise from $R$-parity violation 
and may be attributable to a $Z'$. 

\vspace*{-0.5cm}
\nn
\begin{figure}[htbp]
\centerline{
\psfig{figure=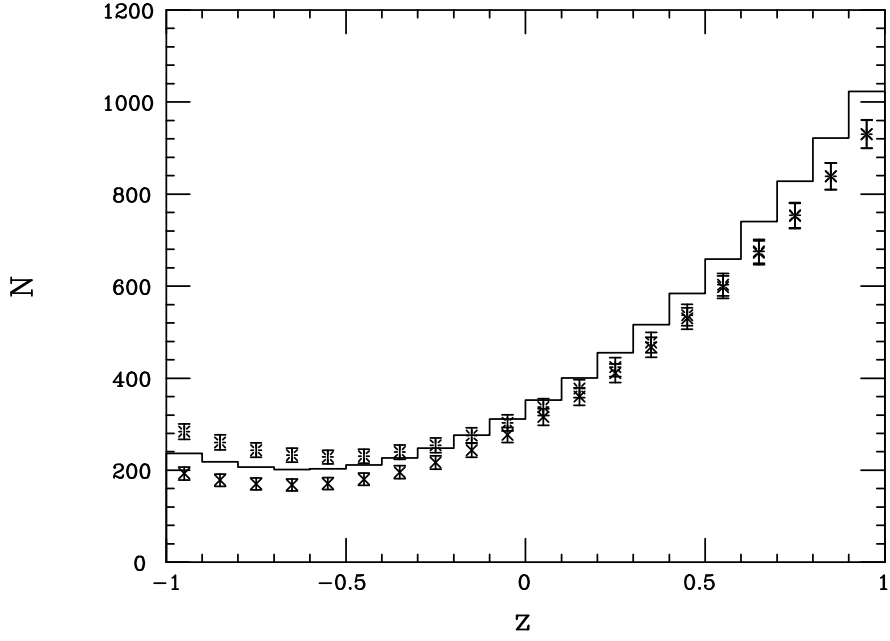,height=10.5cm,width=14cm,angle=-90}}
\vspace*{-10mm}
\centerline{
\psfig{figure=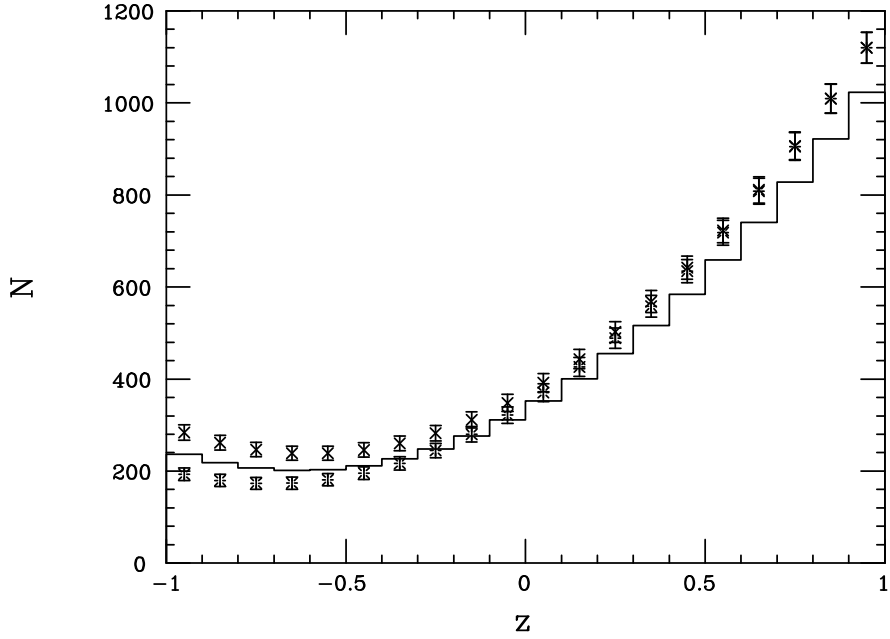,height=10.5cm,width=14cm,angle=-90}}
\vspace*{-0.9cm}
\caption{Same as the previous figure but now including a 3 TeV $Z'$ exchange 
in the $s$-channel. The magnitude of all $Z'$ couplings is taken to be be the 
same value, \ie, $|g^{e\prime,f\prime}_{L,R}|=0.3c$, for purposes of 
demonstration. In the top panel, the relative signs of ($g^{e\prime}_{L}, 
~g^{e\prime}_{R},~g^{f\prime}_{L},~g^{f\prime}_{R}$ are chosen to be 
$(+,-,+,-)[(+,+,+,+)]$ for the upper[lower] series of data points, while in 
the bottom panel they correspond to the choices $(+,-,-,+)[(+,+,-,-)]$ for 
the upper[lower] series, respectively.}
\label{fig3}
\end{figure}
\vspace*{0.4mm}

In the $Z'$ case assuming a fixed gauged boson mass,
 we have four couplings that we 
can freely vary, \ie, $g^{e\prime,f\prime}_{L,R}$. For simplicity 
we will assume 
that all these couplings have the same magnitude (but we strongly emphasize 
that this need not be the case), \ie, $|g^{e\prime,f\prime}_{L,R}|=0.3c$, 
and in 
this case the four possible relative sign combinations can lead to quite 
different angular distributions as shown in Fig.3. Here we see that depending 
on the choice of relative signs, the $Z'$ exchange can lead to positive or 
negative modifications 
in the distribution in both the positive and negative ranges of $z$. Clearly 
if these four couplings were allowed to vary freely almost any reasonable 
shift in the distribution could be obtainable. We would thus expect that some 
choice of $Z'$ couplings could be made to completely simulate the $\tilde \nu$ 
signal. 

\vspace*{-0.5cm}
\nn
\begin{figure}[htbp]
\centerline{
\psfig{figure=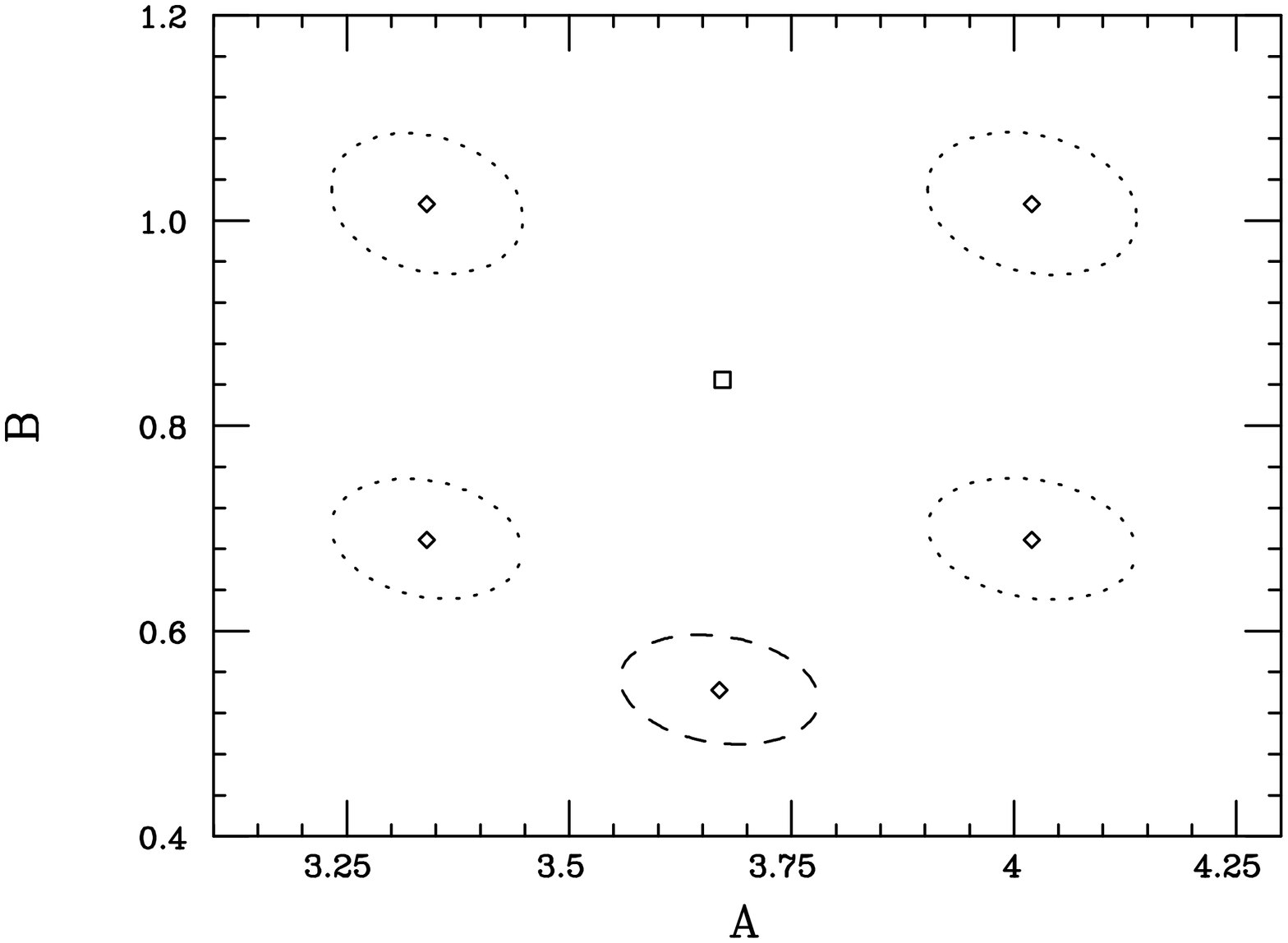,height=14cm,width=17cm,angle=-90}}
\vspace*{-1cm}
\caption{$95\%$ CL fits to the values of $A$ and $B$ for the data generated 
with $\tilde \nu$ exchange(dashed region) and for the data generated for the 
four possible choices of $Z'$ couplings(dots). The SM result is represented 
by the square in the center of the figure while the diamonds are the locations 
of the best fits.}
\label{fig4}
\end{figure}
\vspace*{0.4mm}

How would the analysis then proceed? The exact form of the angular distribution 
given above suggests the following approach: once deviations in the 
distribution are observed a two parameter fit of the data could be performed 
to a trial distribution of the form
\begin{equation}
{d\sigma\over {dz}}\sim  A(1+z)^2+B(1-z)^2\,,
\end{equation}
where from the exact expression above 
we see that $A\sim|f^s_{LR}|^2+|f^s_{RL}|^2$ 
and $B\sim|f^s_{LL}|^2+|f^s_{RR}|^2$. A fit to this distribution may isolate 
whether the new physics occurs in the value of coefficient $A$, $B$, or both. 
In the SM and $Z'$ cases both $A$ and $B$ are 
constants, but $B$ picks up an additional $z$ dependence in the case of 
$\tilde \nu$ exchange. If this additional $z$ dependence is strong, \ie, we 
are not in the contact interaction limit, then the $\chi^2$ of the fit 
assuming a constant $B$ in the case of $\tilde \nu$ exchange 
will be poor. Let us consider the `data' as shown in 
Figs. 2 and 3 as input into this analysis for purposes of demonstration; the 
result of the fitting procedure for these sample cases 
is shown in Fig.4. Here we see that all five 
sets of `data' lie quite a distance from the SM point clearly indicating 
the presence of new physics at a high confidence level. In the 
case of $\tilde \nu$ exchange we see that the value of $A$ arising from the 
fit is in excellent agreement with the expectations of the 
SM, while in the $Z'$ case the values of 
both $A$ and $B$ have been altered. Note that all five allowed regions are 
statistically well separated from each other. Futhermore, in all cases 
the resulting confidence 
level (CL) of the fits are very good indicating no special sensitivity to any 
variation in the value of $B$ with $z$ for $\tilde \nu$ exchange. 
(Numerically, we find the bin-averaged value of $B$ to vary between 0.546 and 
0.518 as we go from large negative to large positive $z$.) Given the 
distribution of the $Z'$ results one can imagine that a suitably chosen 
conspiratorial set of values for the couplings $g^{e,f\prime}_{L,R}$ could 
lead to a substantial overlap with the extracted $\tilde \nu$ coupling region 
in which case the two new physics sources would not be distinguishable. 
Except for this conspiratorial region, however, it would appear that the fits 
to the angular distribution do provide a technique to separate these two SM 
extensions.

\vspace*{-0.5cm}
\nn
\begin{figure}[htbp]
\centerline{
\psfig{figure=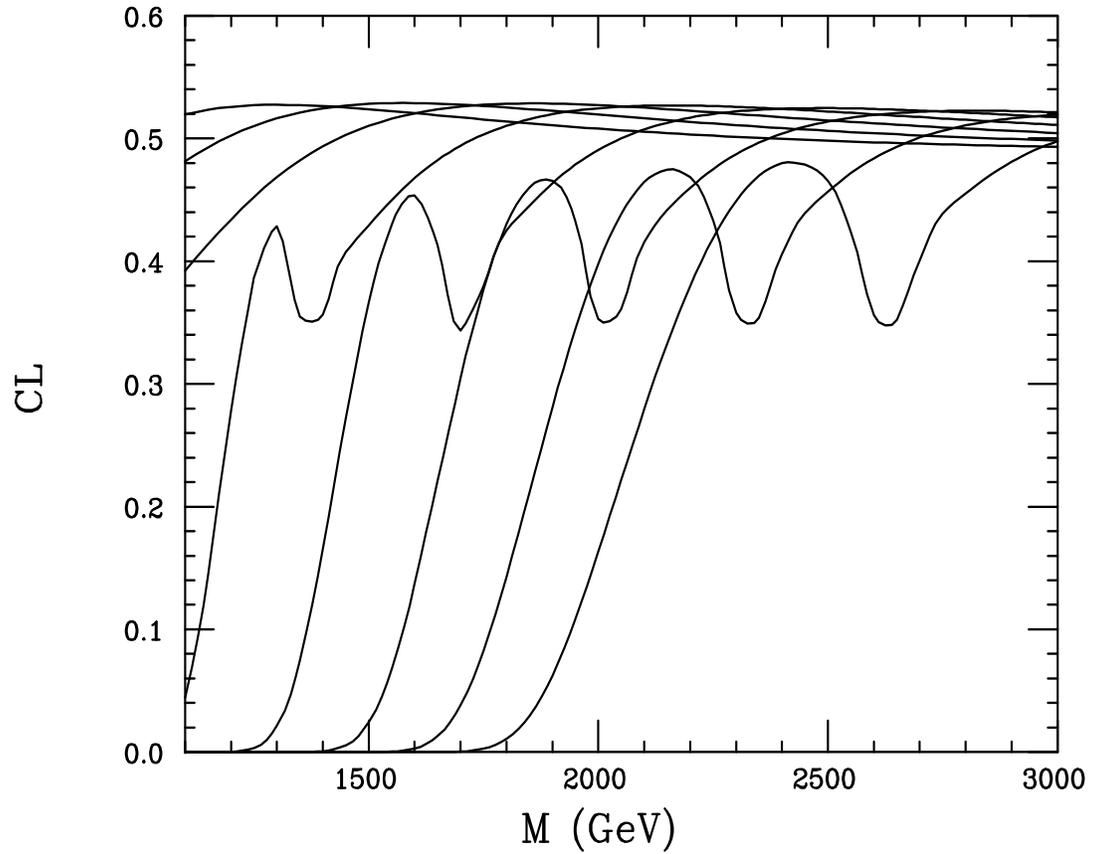,height=14cm,width=17cm,angle=-90}}
\vspace*{-1cm}
\caption{Average confidence level of the best fit to the parameters $A$ and 
$B$ as a function of the $\tilde \nu$ mass in the case of $t$-channel 
$\tilde \nu$ exchange for various values of the Yukawa 
coupling $\lambda$ in the range 0.3 to 1.0 in steps of 0.1 from top left to 
lower right. }
\label{fig5}
\end{figure}
\vspace*{0.4mm}

As discussed above, when the value of $\lambda/m_{\tilde \nu}$ becomes 
sufficiently large it will become apparent that the fit with a constant 
$B$ will no longer provide a good fit.  Exactly when does this happen? To 
address this question we vary both $\lambda$ and the $\tilde \nu$ mass and 
perform a multitude of fits assuming that $A$ and $B$ are constant and 
obtain the confidence level of the best fit for each case. The result of this 
analysis is shown in Fig.5. In this figure  we see that typically one finds 
that this type of fit begins to fail in a qualitative way when 
$\lambda/m_{\tilde \nu}\geq 0.5~TeV^{-1}$. For much smaller values of this 
parameter, as in the sample case above, the data will be insensitive to the 
nature of the $t$-channel exchange and we will be living in the contact 
interaction limit of parameter space. How does this bound scale with the 
collider energy? Since the $t-$channel $\tilde \nu$ exchange interferes 
directly with the SM contribution, assuming that most of the error is 
statistical in origin, we expect the bound on the ratio 
$\lambda/m_{\tilde \nu}$ to roughly scale as $\sim (L\cdot s)^{-1\over 4}$, 
where $L$ in the integrated luminosity and $s$ in the machine center of mass 
energy.

\section{$s$-channel $\tilde \nu$ Exchange}

When a $Z'$ or $\tilde \nu$ are exchanged in the $s$-channel, 
the general form of 
the cross section with a polarized electron beam can be written as: 
\begin{eqnarray}
{d\sigma\over {dz}} &=& {\pi\alpha^2\over {8s}}\biggl[(1+z)^2
\left\{{1+P\over {2}}|f^s_{LR}|^2 +{1-P\over {2}}|f^s_{RL}
|^2\right\}\nonumber \\
&+&(1-z)^2\left\{{1+P\over {2}}|f^s_{LL}|^2+{1-P\over {2}}
|f^s_{RR}|^2\right\}+4\left\{{1+P\over {2}}|f^t_{LL}|^2+{1-P\over {2}}
|f^t_{RR}|^2\right\}\biggr]
\,,
\end{eqnarray}
where $f^s_{ij}$ are obtainable above and 
\begin{eqnarray}
f^t_{LL}&=&f^t_{RR}=0~~~~~~~~~~(Z^\prime)\,, \nonumber \\
f^t_{LL}&=&f^t_{RR}={1\over 2}C_{\tilde \nu}P^s_{\tilde \nu}~~~~(\tilde \nu)\,,
\end{eqnarray}
with $P^s_{\tilde \nu}=s/(s-m_{\tilde \nu}^2+im_{\tilde \nu}\Gamma_{\tilde \nu})
\simeq s/(s-m_{\tilde \nu}^2)$ in the same limit as employed above. Our first 
step here is to determine the search reach for a $\tilde \nu$ being 
exchanged in the 
$s$-channel. Our standard analysis yields the results shown in Fig.6; note that 
the search reach for a fixed value of $\lambda_0$ is somewhat larger in the 
$t$-channel than in the $s$-channel but generally comparable in magnitude. 
Note that here $\lambda_0^2=\lambda_1\lambda_2$, with $\lambda_{e,f}$ being the 
values of the Yukawa couplings for the $\tilde \nu$ to initial state 
electrons and the fermion $f$ in the final state.

\vspace*{-0.5cm}
\nn
\begin{figure}[htbp]
\centerline{
\psfig{figure=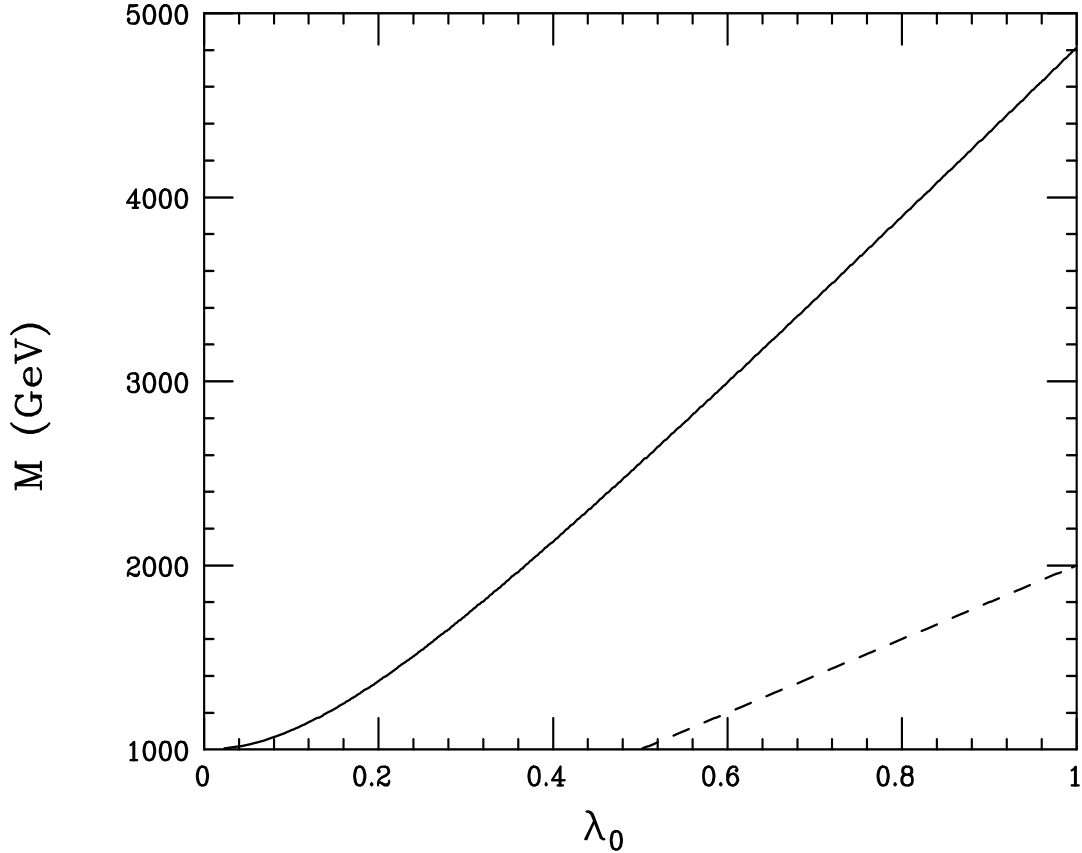,height=14cm,width=17cm,angle=-90}}
\vspace*{-1cm}
\caption{Same as Fig. 1 but now for $s$-channel $\tilde \nu$ exchange. Here 
$\lambda_0^2$ equals the product of the relevant Yukawa couplings in the 
superpotential. The typical region excluded by low energy data is that below 
the dashed curve in the lower right hand corner.}
\label{fig6}
\end{figure}
\vspace*{0.4mm}

As before a short analysis demonstrates that single beam polarization will not 
help distinguish these two 
new physics models due the small value of the resulting 
asymmetry, so we set $P=0$ and again examine the angular distribution. First, 
we note that when a $\tilde \nu$ is exchanged in the $s$-channel the 
angular distribution picks up a constant, \ie, $z$-independent term: 
\begin{equation}
{d\sigma\over {dz}}\sim  A(1+z)^2+B(1-z)^2+C\,,
\end{equation}
with $A,B$ given as before and here $C\sim 2[C_{\tilde \nu}P_{\tilde \nu}]^2$. 
As expected, when the value of the constant $C$ is sufficiently 
large it will become apparent that the resulting fit which assumes that 
only $A$ and $B$ are present is no 
longer valid due to an increase in $\chi^2$ and a lower confidence level. 
However, for moderate coupling strengths we find that it is possible to 
adjust the values of $A$ and $B$ to mask the contributions of the $C$ term. 
In Fig.7 we show the CL obtained by performing a large number of fits to the 
parameters $A$ and $B$ for different values of both $\lambda_0$ and the 
$\tilde \nu$ mass from generating `data' samples via Monte Carlo. 
For small $\lambda_0$'s or large masses, as in the above 
example, we see that the CL of the fit is always quite good. In the opposite 
limit, the fit fails and the CL is quite small. 
Typically, we see that the fit begins 
to fail qualitatively when $\lambda_0/m_{\tilde \nu}\geq 0.25-0.30~TeV^{-1}$. 
This reach in coupling-mass parameter space is not very good and so we seek 
other observables with which to extend our reach.

\vspace*{-0.5cm}
\nn
\begin{figure}[htbp]
\centerline{
\psfig{figure=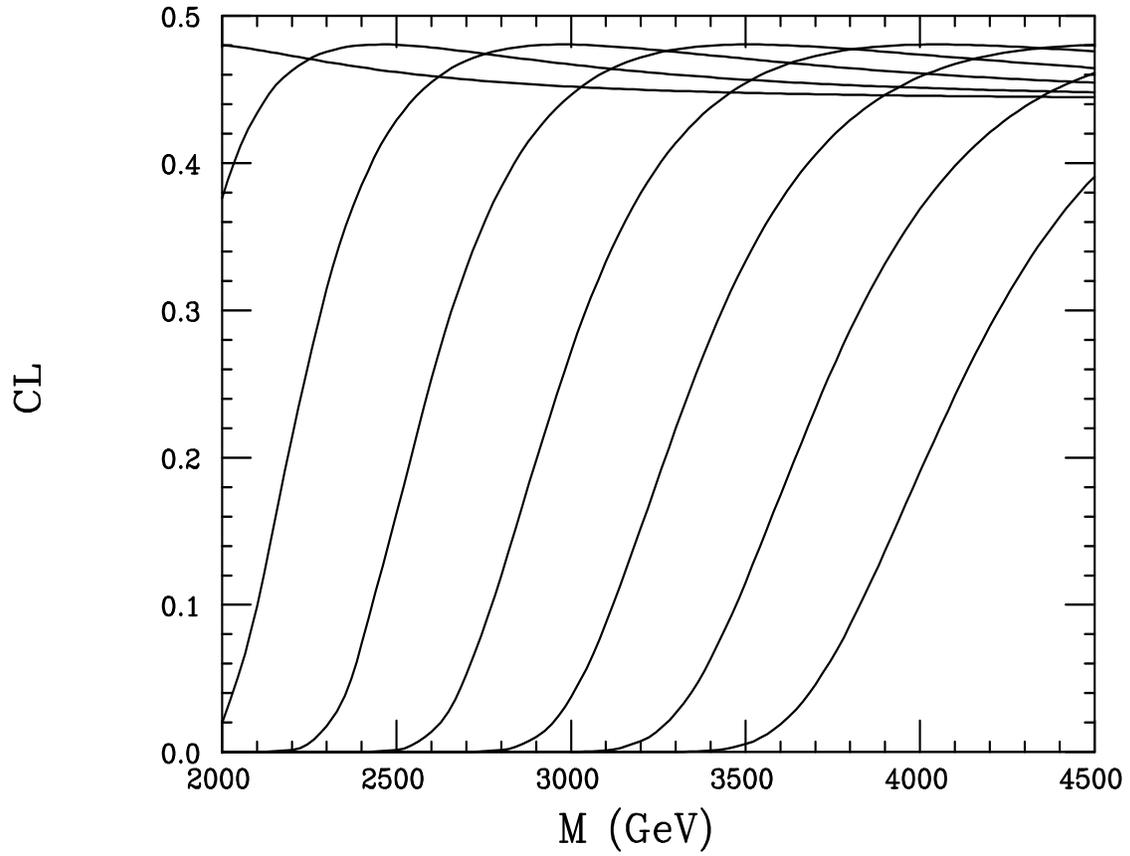,height=14cm,width=17cm,angle=-90}}
\vspace*{-1cm}
\caption{Average confidence level of the best fit to the parameters $A$ and 
$B$ as a function of the $\tilde \nu$ mass in the case of $s$-channel 
$\tilde \nu$ exchange for various values of the Yukawa 
coupling $\lambda_0$ in the range 0.3 to 1.0 in steps of 0.1 from top left to 
lower right. }
\label{fig7}
\end{figure}
\vspace*{0.4mm}

In the case where a $\tau$ pair is being produced in the final state we can 
employ a clever idea used by Bar-Shalom, Eilam and Soni(BES){\cite {bumps}} 
in a somewhat different context. If the $\tau$ spins can be analyzed, a 
spin-spin correlation can be formed which is sensitive to the spin of any new 
particle exchanged in the $s$-channel. Integrating over all production angles, 
this quantity can be written as an asymmetry:
\begin{equation}
B_{zz}={|f^s_{LR}|^2+|f^s_{LR}|^2+|f^s_{LR}|^2+|f^s_{LR}|^2-
{3\over {4}}(|f^t_{LL}|^2+|f^t_{RR}|^2)\over {|f^s_{LR}|^2+|f^s_{LR}|^2+
|f^s_{LR}|^2+|f^s_{LR}|^2+{3\over {4}}(|f^t_{LL}|^2+|f^t_{RR}|^2)}}\,,
\end{equation}
where we see immediately that for the case of the SM or a $Z'$ one obtains 
$B_{zz}=1$ whereas a $\tilde \nu$ exchange in the $s$-channel will force this 
observable to smaller, even negative values. In Fig.8 we display the value of 
the asymmetry $B_{zz}$ as a function of the $\tilde \nu$ mass for several 
values of $\lambda_0$. Even if the efficiency for making this spin-spin 
correlation measurement is only $50\%$, the anticipated statistical error on 
this quantity will be of order $1\%$ since there are about 9000 $\tau$-pairs 
in the data sample. Thus a value of $B_{zz}$ below 
$\simeq 0.95$ would provide a very strong indication that there is a 
scalar exchange in the $s$-channel. From the figure we see that this implies 
that the parameter space region 
$\lambda_0/m_{\tilde \nu}\geq 0.15-0.20~TeV^{-1}$ would certainly be probed by 
such measurements. Unfortunately, this technique does not help us in the case 
of a corresponding $t$-channel exchange. 

\vspace*{-0.5cm}
\nn
\begin{figure}[htbp]
\centerline{
\psfig{figure=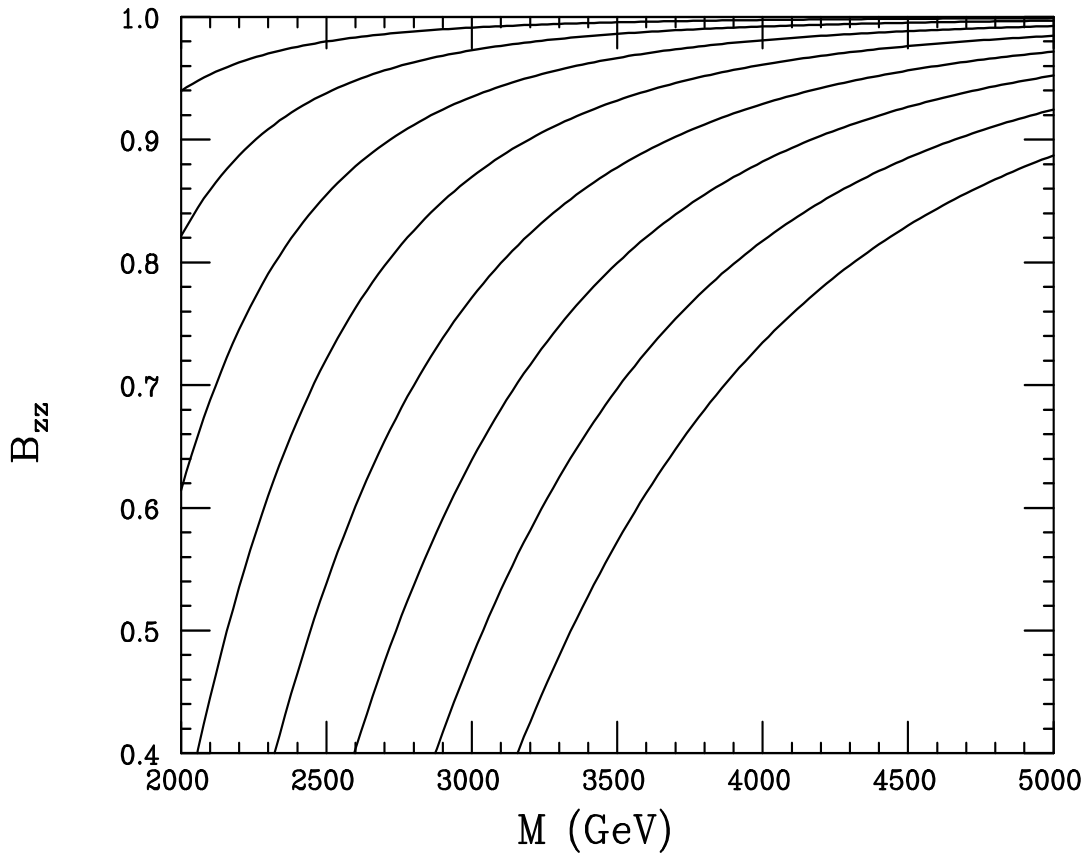,height=14cm,width=17cm,angle=-90}}
\vspace*{-1cm}
\caption{Double $\tau$ spin asymmetry at a 1 TeV NLC as a function of the 
$\tilde \nu$ mass for different values of the Yukawa coupling $\lambda_0$. 
From left to right, $\lambda_0$ varies from 0.3 to 1.0 in steps of 0.1 as in 
the previous figure. In the case of either the SM or a $Z'$, $B_{zz}=1$.}
\label{fig8}
\end{figure}
\vspace*{0.4mm}

It is apparent that for non-$\tau$ pair final states we cannot use this trick. 
While we have already observed that single beam polarization is not useful, if 
{\it both} initial beams can be polarized{\cite {zdr}} more observables can 
be investigated. In this case, 
integration over $z$ gives the following expression for the cross section 
with two polarized beams: 
\begin{equation}
\sigma(P_1,P_2)\sim [LL]\{|f^s_{LR}|^2+|f^s_{LL}|^2\}+[RR]\{|f^s_{RL}|^2
+|f^s_{RR}|^2\}+{3\over {4}}[LR]\{|f^t_{LL}|^2+|f^t_{RR}|^2\}\,,
\end{equation}
where we have employed the notation {\cite {cuypers}}, 
\begin{eqnarray}
{[LL]} &=& {1\over {4}}[1+P_1+P_2+P_1P_2]\,, \nonumber \\
{[RR]} &=& {1\over {4}}[1-P_1-P_2+P_1P_2]\,, \nonumber \\
{[LR]} &=& {1\over {2}}[1-P_1P_2]\,,
\end{eqnarray}
with $P_{1,2}$ being the polarizations of the incoming electron and positron 
beam respectively. From these cross sections a double polarization asymmetry 
can be obtained:
\begin{equation}
A_{double}={\sigma(+,+)+\sigma(-,-)-\sigma(-,+)-\sigma(+,-)\over {\sigma(+,+)
+\sigma(-,-)+\sigma(-,+)+\sigma(+,-)}}\,. 
\end{equation}
Let us assume that $P_1=P_{e^-}=0.90$ while $P_2=P_{e^+}=0.65$ as given in 
Ref.{\cite {zdr}}; we then calculate $A_{double}$ readily and obtain 
a value of 0.585 for both the SM and when a $Z'$ is present. However, as in 
the case of $B_{zz}$, the presence of $\tilde \nu$ exchange in the $s$-channel 
can lead to significantly smaller values of $A_{double}$. It is interesting 
to note that this double polarization asymmetry would not have helped in the 
case of $t$-channel $\tilde \nu$ exchange since it and the $Z'$ contribute 
to the same amplitudes. 

\vspace*{-0.5cm}
\nn
\begin{figure}[htbp]
\centerline{
\psfig{figure=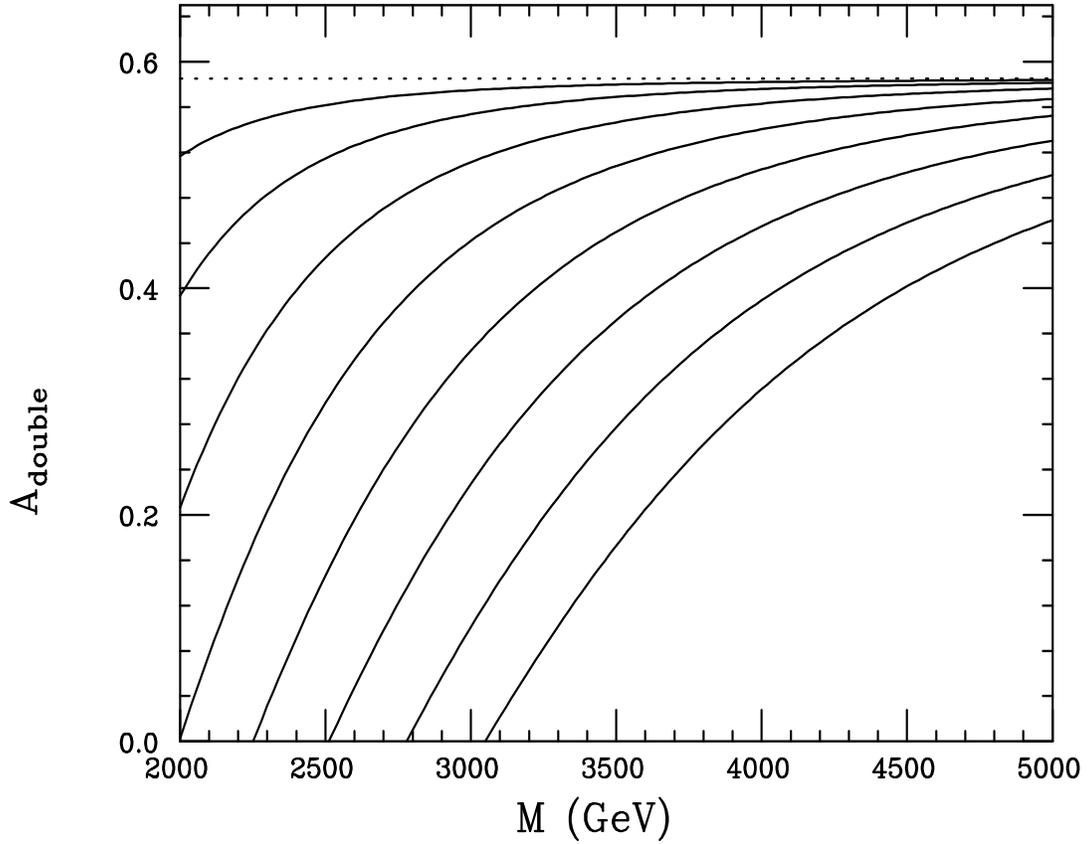,height=14cm,width=17cm,angle=-90}}
\vspace*{-1cm}
\caption{The double polarization asymmetry, $A_{double}$ as a function of the 
$\tilde \nu$ mass at a 1 TeV NLC for different choices of $\lambda_0$. From  
left to right, $\lambda_0$ varies from 0.3 to 1.0 in steps of 0.1. 
The dotted curve corresponds to 
the value obtained for both the SM and in the case of a $Z'$. }
\label{fig9}
\end{figure}
\vspace*{0.4mm}

Fig.9 shows the set of results obtained for $A_{double}$ in this case as a 
function of the mass of the $\tilde \nu$ assuming various values for the 
Yukawa coupling $\lambda_0$. Since the statistical error on $A_{double}$ is 
again expected to be somewhat less than $1\%$ for our assumed integrated 
luminosity (a value of $0.86\%$ is obtained for the SM) 
it is clear that for values of 
$\lambda_0/m_{\tilde \nu}\geq 0.15-0.20~TeV^{-1}$, a 
statistically significant signal for scalar $s$-channel exchange will be 
observed. This reach is quite comparable to that obtained using the double spin 
asymmetry technique discussed above and is superior to that found by 
an examination of the angular distribution alone. 
How does this bound scale with the 
collider energy? Since the $s-$channel $\tilde \nu$ exchange does not directly 
interfere with the SM contribution, assuming that most of the error is 
statistical in origin, we expect the bound on the ratio 
$\lambda/m_{\tilde \nu}$ to roughly scale as $\sim (L\cdot s^3)^{-1\over 8}$, 
where $L$ in the integrated luminosity and $s$ in the machine center of mass 
energy.

\section{Bhabha Scattering}

Bhabha scattering represents the most difficult case of the ones we have 
considered since $\gamma$ and $Z$ exchanges are already present in both the 
$s$- and $t$-channels in the SM 
and in fact the $t$-channel $\gamma$ pole dominates. Allowing for $s$- and 
$t$-channel $Z'$ or $\tilde \nu$ exchange for the case where both 
electron and positron beams are polarized, the differential cross section 
can be written as 
\begin{eqnarray}
{d\sigma\over {dz}} &=& {\pi\alpha^2\over {8s}}\biggl[(1+z)^2
\left\{[LL]|f^s_{LR}|^2 +[RR]|f^s_{RL}|^2+[LL]|f^t_{LR}|^2 
+[RR]|f^t_{RL}|^2\right. \nonumber \\
&+&\left. 2[LL]f^s_{LR}f^t_{LR}+2[RR]f^s_{RL}f^t_{RL}\right\}\nonumber \\
&+&(1-z)^2\left\{[LL]|f^s_{LL}|^2+[RR]|f^s_{RR}|^2\right\}+
2[LR]\left\{|f^t_{LL}|^2+|f^t_{RR}|^2\right\}\biggr]\,,
\end{eqnarray}
where the $f^s_{ij}$ can be obtained from the expressions above and 
\begin{eqnarray}
f^t_{LR}&=&{s\over {t}}+P^t_Z(g^e_L)^2\oplus P^t_{Z'}(g^{e\prime}_L)^2\oplus 0
\,, \nonumber \\
f^t_{RL}&=&{s\over {t}}+P^t_Z(g^e_R)^2\oplus P^t_{Z'}(g^{e\prime}_R)^2\oplus 0
\,, \nonumber \\
f^t_{LL}&=&{s\over {t}}+P^t_Zg^e_Lg^e_R\oplus P^t_{Z'}g^{e\prime}_Lg
^{e\prime}_R\oplus {1\over {2}}C_{\tilde \nu}P_{\tilde \nu}\,, \nonumber \\
f^t_{RR}&=&{s\over {t}}+P^t_Zg^e_Lg^e_R\oplus P^t_{Z'}g^{e\prime}_Lg
^{e\prime}_R\oplus {1\over {2}}C_{\tilde \nu}P_{\tilde \nu}\,,
\end{eqnarray}
with $P^t_{Z,Z'}=s/(t-M_{Z,Z'}^2)$. The search reaches for a $Z'$ or 
$\tilde \nu$ in this channel are found to be very comparable to that of the 
case of $s$-channel exchange discussed above.

\vspace*{-0.5cm}
\nn
\begin{figure}[htbp]
\centerline{
\psfig{figure=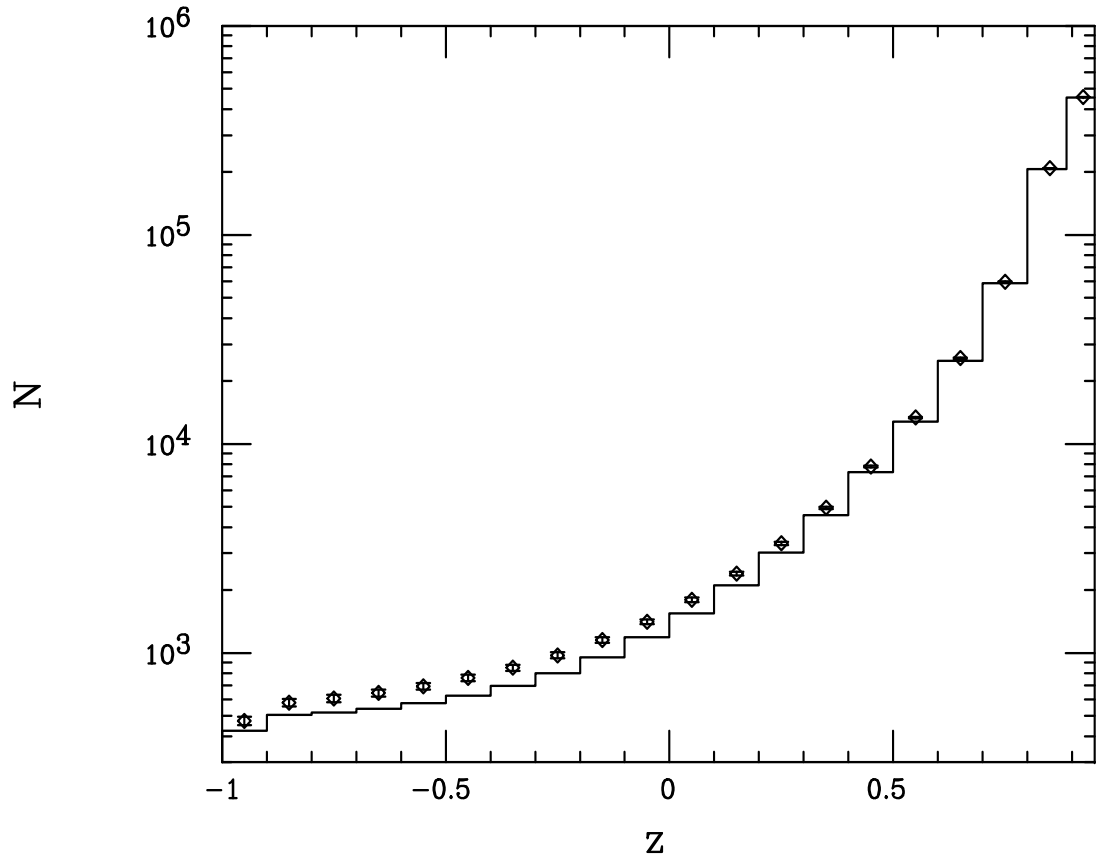,height=14cm,width=17cm,angle=-90}}
\vspace*{-1cm}
\caption{Same as Fig.2 but now for the case of Bhabha scattering. Angular cuts 
as described in the text have been employed to render the cross section 
finite in the forward direction.}
\label{fig10}
\end{figure}
\vspace*{0.4mm}

To examine this cross section in any 
detail, angular cuts are necessary due to the photon pole in the forward 
direction. We first employ a weak cut of $|z|<0.985$, corresponding to 
$\theta \geq 10^o$, which is motivated 
by detector requirements{\cite {zdr}}. This has little effect in the backward 
direction and leaves an enormous rate in the forward direction. 
To further tame the cross section in this direction we strengthen this cut 
to $z<0.95$ to remove more of the photon pole. The result of this procedure 
for the SM and for the case of a 3 TeV $\tilde \nu$ with $\lambda=0.5$ is shown 
in Fig.10 for a $\sqrt s=1$ TeV NLC assuming unpolarized beams and 
an integrated luminosity of 
$150~fb^{-1}$. As one might expect, the distribution in the far forward 
direction is overwhelmingly 
dominated by the photon pole and hence there is no signal for new 
physics there even with the large statistics available.  In the 
backwards direction, the $\tilde \nu$ exchange is seen to lead to a 
characteristic and statistically significant increase in the cross section 
above that predicted by the SM. Since $\tilde \nu$ exchange can only increase 
the cross section in the backward region, any observed 
decrease in the cross section may be attributable to a $Z'$. As can be seen 
in Fig.11, when the product of $Z'$ 
couplings $g^{e\prime}_Lg^{e\prime}_R>(<)0$, 
the resulting cross section is seen to increase(decrease) in this case.

\vspace*{-0.5cm}
\nn
\begin{figure}[htbp]
\centerline{
\psfig{figure=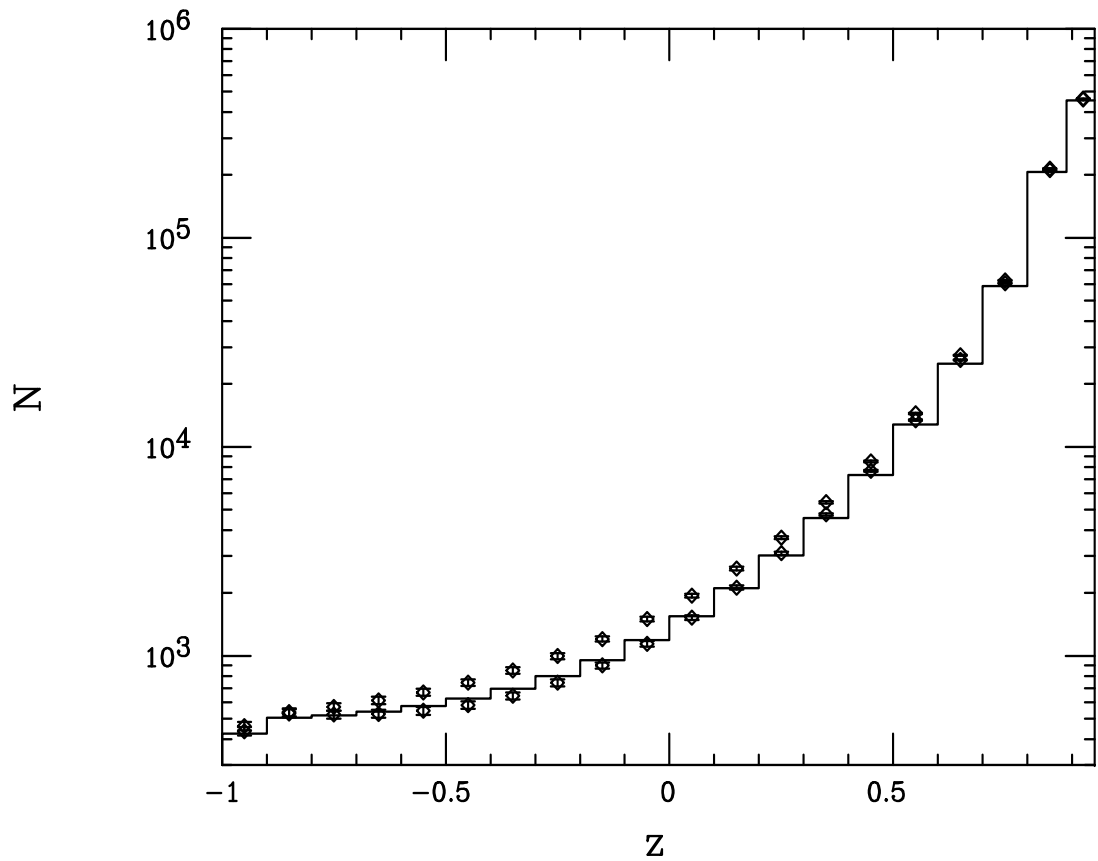,height=14cm,width=17cm,angle=-90}}
\vspace*{-1cm}
\caption{Same as the previous figure but now for a 3 TeV $Z'$ in comparison 
to the SM. The upper(lower) set of data points corresponds to 
$g^{e\prime}_L=g^{e\prime}_R=0.5c$($g^{e\prime}_L=-g^{e\prime}_R=0.5c$).}
\label{fig11}
\end{figure}
\vspace*{0.4mm}

From this discussion it is clear that using the Bhabha scattering angular 
distribution alone it will be possible to easily distinguish new physics in 
the form of a $\tilde \nu$ from a $Z'$ when $g^{e\prime}_Lg^{e\prime}_R<0$. 
When the product of $Z'$ couplings have the opposite sign we need to use an 
additional observable. One immediate possibility is to employ $A_{double}$ as 
defined above in the case that both initial 
beams are polarized. However, due to the 
dominance of the photon pole at $z=1$ we limit ourselves to events with $z<0$; 
for the SM this corresponds to about 7000 events when the integrated 
luminosity is 150 $fb^{-1}$ at a 1 TeV NLC after ISR and gives 
$A_{double}(SM)=-0.273\pm 0.011$, obtained 
by taking $P_1=0.90$ and $P_2=0.65$ as in the discussion above. 
A scan of the $\lambda$ 
and $\tilde \nu$ mass parameter space leads us to the observation that 
$\tilde \nu$ exchange always decreases the value of the asymmetry from that 
obtained in the case of the SM.  $Z'$ 
exchange also modifies the value of this asymmetry; unfortunately we find that 
for $g^{e\prime}_Lg^{e\prime}_R>0$, $A_{double}$ also decreases as it does for 
the case of $\tilde \nu$ exchange. Thus $A_{double}$ does not help us resolve  
this potential ambiguity in the case of Bhabha scattering.

\section{Discussion and Conclusion}

In this paper we have considered the problem of how to distinguish two 
potential new physics scenarios from each other below the threshold for direct 
production of new particles at the NLC: $R$-parity violation and a extension 
of the SM gauge group by an additional $U(1)$ factor. Both kinds of new 
physics can lead to qualitatively similar alterations in SM cross sections,  
angular distributions and various asymmetries but differ in detail. These 
detailed differences provide the key to the two major weapons that are useful 
in accomplishing our task: ($i$) the angular distribution of the final state 
fermion and ($ii$) an asymmetry formed by polarizing both beams in the initial 
state, $A_{double}$. The traditional asymmetry, $A_{LR}$, formed when only a 
single beam is polarized, was shown not to be useful for the case of purely 
leptonic processes we considered, but will be useful in an extension of the 
analysis to hadronic final states. This same analysis employed above can be 
easily extended to other new physics scenarios which involve the exchange on 
new particles{\cite {hewett}} as in the case of massive graviton exchange in 
theories with compactified dimensions.

\noindent{\Large\bf Acknowledgements}

The author would like to thank J.L. Hewett, S. Bar-Shalom and members of the 
Future $e^+e^-$ Linear Collider New Phenomena working group for discussions 
related to this analysis. The author would also like to thank J. Conway and 
the Rutgers Physics Department for their hospitality.

\newpage

%
\def\MPL #1 #2 #3 {Mod. Phys. Lett. {\bf#1},\ #2 (#3)}
\def\NPB #1 #2 #3 {Nucl. Phys. {\bf#1},\ #2 (#3)}
\def\PLB #1 #2 #3 {Phys. Lett. {\bf#1},\ #2 (#3)}
\def\PR #1 #2 #3 {Phys. Rep. {\bf#1},\ #2 (#3)}
\def\PRD #1 #2 #3 {Phys. Rev. {\bf#1},\ #2 (#3)}
\def\PRL #1 #2 #3 {Phys. Rev. Lett. {\bf#1},\ #2 (#3)}
\def\RMP #1 #2 #3 {Rev. Mod. Phys. {\bf#1},\ #2 (#3)}
\def\ZPC #1 #2 #3 {Z. Phys. {\bf#1},\ #2 (#3)}
\def\IJMP #1 #2 #3 {Int. J. Mod. Phys. {\bf#1},\ #2 (#3)}

\end{document}